\documentclass[a4paper]{article}
\usepackage{onecolpceurws}
\usepackage{listings}
\usepackage{graphicx}
\usepackage{epsfig}
\usepackage{subcaption}
\usepackage{color}
\usepackage{wrapfig}
\usepackage[utf8]{inputenc}

\title{The Fulib Solution to the TTC 2020 Migration Case}

\author{
Sebastian Copei, sco@uni-kassel.de \and 
Albert Z\"undorf, zuendorf@uni-kassel.de
}

\institution{Kassel University}

\begin{document}


\maketitle

\section{Introduction}

At Kassel University we are working on a solution for bidirectional 
transformations based on event sourcing 
for about a year, now. It turned out, that the TTC 2020 migration case 
\cite{ttc2020migration} is a special case of a bidirectional 
transformation and that our approach provides a reasonable solution for it.

\section{Design}

\begin{figure}[ht] \centering
	\includegraphics[width=0.7\linewidth]{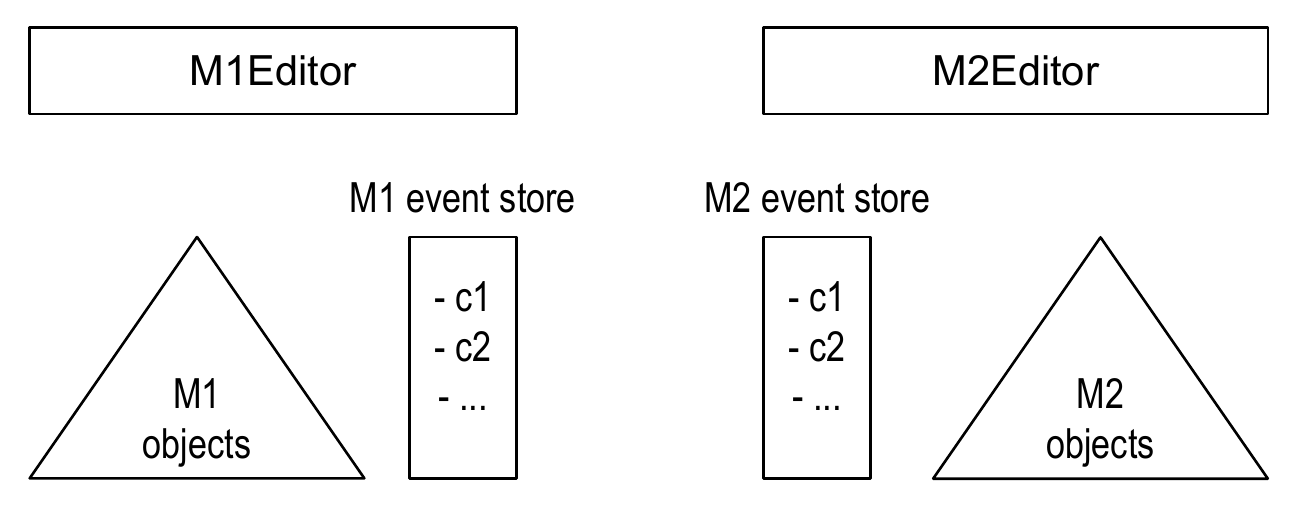}
 \caption{Design}
 \label{fig:design}
\end{figure}

\begin{figure}[ht] \centering
	\includegraphics[width=0.35\linewidth]{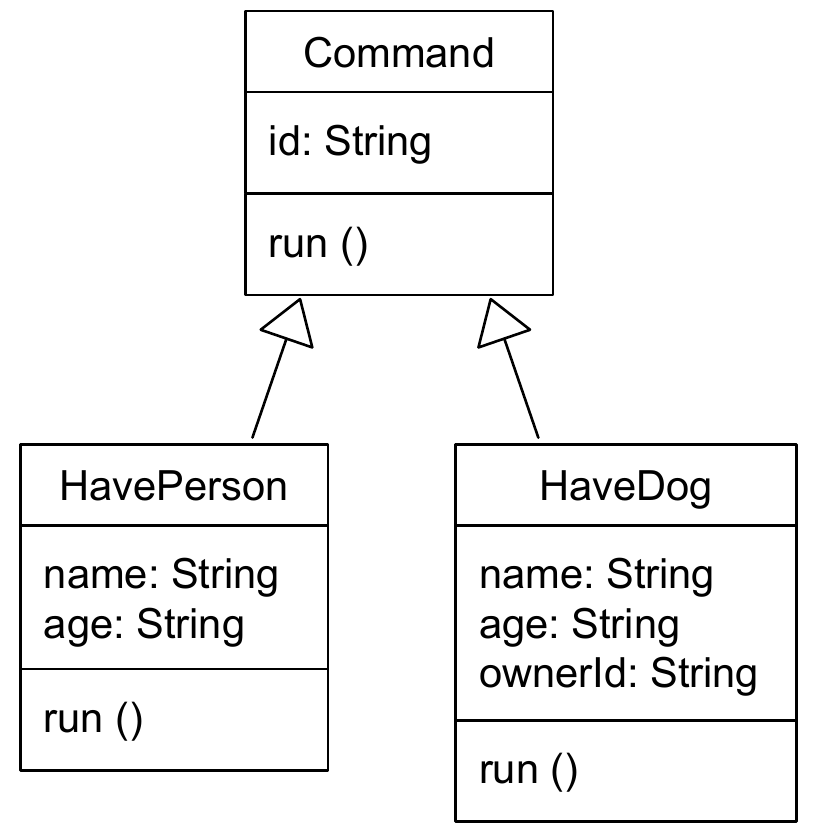}
 \caption{Command Classes}
 \label{fig:classes}
\end{figure}

\begin{figure}[ht] \centering
	\includegraphics[width=0.9\linewidth]{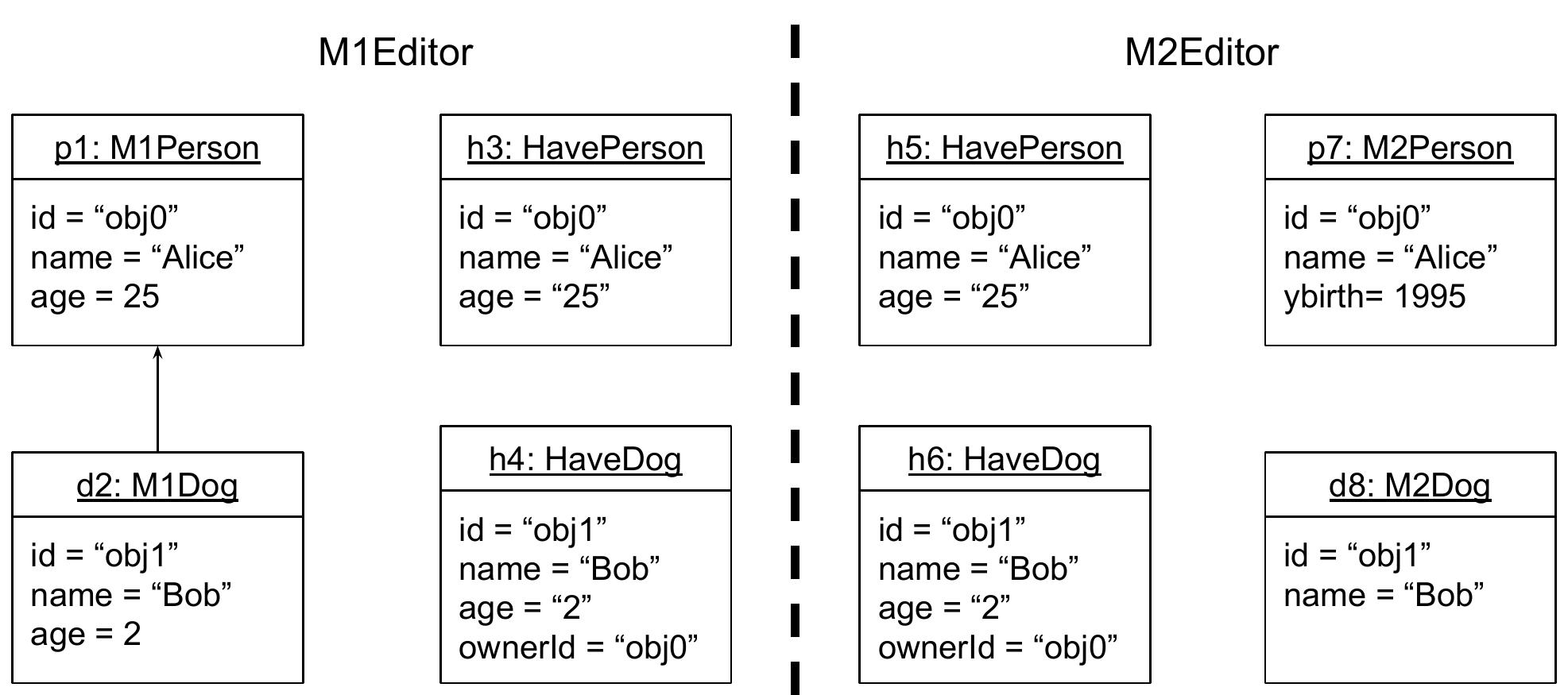}
 \caption{Objects}
 \label{fig:objects}
\end{figure}

The idea and design for our solution stems from Domain Driven Design \cite{evans2004domain} 
and Event Sourcing \cite{vernon2013implementing}. Basically, we use two editors 
\textit{M1Editor} and \textit{M2Editor}, one for each model of the case study, 
cf. Figure~\ref{fig:design}.
Each editor holds the current object model (based on the corresponding ecore model). In addition, 
each editor provides editing commands following the command design pattern of 
\cite{gamma1993design}.  All operations on the object model are encapsulated within 
editor commands. Each editor keeps track of all executed commands (and the used command 
parameters) within its \textit{event store}, cf. Figure~\ref{fig:design}. 
To enable collaboration of M1Editor and M2Editor, both editors provide the same set of commands: 
a \textit{HavePerson} command with parameters \textit{id}, \textit{name}, and \textit{age} 
and a \textit{HaveDog} command with parameters \textit{id}, \textit{ownerId}, \textit{name}, 
and \textit{age}, cf. Figure~\ref{fig:classes} and Figure~\ref{fig:objects}. 

While both editors provide the same commands (with the same parameters), 
each editor implements
the command execution differently according to its specific ecore model. 
As an example, Listing~\ref{m2.run} shows the implementation of the \textit{HavePerson} 
command within M2Editor. Line~15 of Listing~\ref{m2.run} 
shows how the age parameter 
of the HavePerson command is turned into a \textit{ybirth} value for model M2. 

Generally, we consider M1Editor and M2Editor as independent programs 
that may run on different computers, concurrently. 
Therefore each editor is able to serialize its event store (in yaml format) 
and to send its commands to the other editor. Correspondingly, 
the receiving editor is able to deserialize and execute the commands, too. 

In \cite{copei18491mx} we have developed theoretical foundations for this command (or event) 
sharing between multiple editors: basically, we require that multiple commands that address 
the same object (have the same id) overwrite each other, i.e.: if c1 and c2 are commands 
with the same id than applying c1 and then c2 is similar to applying only c2. In addition, 
commands that work on different objects may be executed in any order (are commutative), 
i.e. if c3 and c4 are commands with different ids, applying first c3 and then c4 must result 
in the same object model as applying c4 first and c3 second. While these are pretty strong 
conditions, it turned out to be easy to implement the commands according to these rules. 
For the TTC2020 migration case we will discuss this in Section~\ref{sec:commands}. 

Provided with overwriting and commutative commands, our editors are able to merge commands 
executed by themselves or received from another editor: if a new command arrives that uses
an id that is already used by some old command, the new command is executed and then the old 
command is replaced by the new command in the event store. If a new command arrives 
that uses a new id, the new command is executed and added to the event store. The event store 
is treated as a set, i.e. the order of the commands (and the order of the command execution) 
does not matter due to our commutativity condition. 

Overall this editor design enables us to implement the TTC2020 migration case as outlined 
in Figure~\ref{fig:migration}. The migrate step is invoked by the test or benchmark Tasks. 
The loading of objects is done by loading the corresponding 
xml files via EMF mechanisms. The parsing steps are discussed in Section~\ref{sec:parsing}.
The sending is done via our yaml serialization and deserialization. The execution 
of commands is discussed in Section~\ref{sec:commands}. The modification step is again 
done by the test or benchmark tasks. Similarly, the test and benchmark tasks invoke the 
migrate back step. The migrate back step uses parsing and sending and execution similar 
to the migrate forward step. However, some details dealing with missing information are discussed 
in Section~\ref{sec:parsing}.

\begin{figure}[ht]  
\begin{enumerate}
\item migrate M1 to M2
    \begin{enumerate}
    \item load M1 objects
    \item parse M1 objects into M1 event store
    \item send M1 commands to M2Editor 
    \item execute commands by M2Editor creating M2 objects and filling M2 event store
    \end{enumerate}
\item modify M2 objects to M2'
\item migrate M2' back to M1' 
    \begin{enumerate}
    \item parse (modified) M2' objects and detect new commands and merge the new commands into
          the old M2 event store
    \item send the updated M2 commands to the M1Editor
    \item execute (modified and new) commands on M1 objects and update M1 event store
    \end{enumerate} 
\end{enumerate}
\caption{The Fulib Migration Approach} \label{fig:migration}
\end{figure}

\section{Commands}\label{sec:commands}

Our approach relies on commands that are shared between the two editors. 
In the simple TTC20 Migration case we need only two commands, 
cf. Figure~\ref{fig:classes} and Figure~\ref{fig:objects}. To facilitate
the merging of concurrent edits, we require that commands are overwriting 
and commutative. Thus, if we e.g. execute a \textit{HavePerson} command 
with a certain id on an empty model, the first time we shall create a 
Person object and fill its attributes. However the second time, we shall 
not create a second Person object but we shall lookup the 
already existing object and just adjust its attributes. In Line~6 of
Listing~\ref{m2.run} our \textit{HavePerson} command achieves this 
behavior by using the \textit{getOrCreatePerson} method of its editor. 
Our editors have hash tables for each model class (i.e. for Person and Dog)
and the \textit{getOrCreatePerson} method looks up this hash table. If the hash 
table already has a Person with the given id, this Person is returned. 
Otherwise, \textit{getOrCreatePerson} creates a Person object, initializes its id, 
adds it to the hash table, and then returns the new Person. 
Thus, you may call \textit{getOrCreatePerson} with a certain id as often as you
like, it will always return the same Person object.

\begin{lstlisting}[language=Java, numbers=left, captionpos=b, 
label={m2.run}, escapeinside={\%}{\%},
caption={M2.HavePerson::run()}
]
public class HavePerson extends Command
{
   @Override
   public Object run(M2Editor editor)
   {
      EObject person = editor.getOrCreatePerson(getId());
      EClass personClass = (EClass) editor.getEmfModel().getEClassifier("Person");
      person.eSet(personClass.getEStructuralFeature("name"), name);
      EStructuralFeature ageFeature = personClass.getEStructuralFeature("age");
      if (ageFeature != null) {
         person.eSet(ageFeature, this.age);
      }
      EStructuralFeature ybirthFeature = personClass.getEStructuralFeature("ybirth");
      if (ybirthFeature != null) {
         person.eSet(ybirthFeature, 2020 - this.age);
      }

      return person;
   }
   ...
\end{lstlisting}

This getOrCreate mechanism also helps us to achieve commutativity for our 
commands. Commutativity for commands requires e.g. that you can execute a 
HaveDog command before you execute the HavePerson command for the dog 
owner. To allow this, our HaveDog commands uses getOrCreatePerson with 
the ownerId to retrieve the corresponding Person object. If the owner already
exists, we just use it, otherwise, the owner object is created on the fly. 
In the latter case, a subsequent execution of the HavePerson command with the 
same id will retrieve the already existing Person object and then fill the 
Person's name and age (or ybirth). 

Once the targeted Person object has been retrieved, Line~7 to Line~16
of Listing~\ref{m2.run} fill the attributes of that Person. 
In this solution, we use EMF dynamic editing features to set the name 
and age or ybirth attributes. The different migration tasks use different 
ecore models with different properties. We store the current ecore model 
within the editors and thus the commands just query the current ecore model 
e.g. whether the current Person has either an age attribute or an ybirth 
attribute. Thus one implementation of our M2.HavePerson command suffices 
to address all different migration tasks of the TTC2020 migration case. 
Note, the M1.HavePerson command also has only one implementation for all 
tasks, however this implementation is slightly simpler as the one
of M2.HavePerson as no variant of model M1 deals with ybirth attributes.

\section{Parsing} \label{sec:parsing}

The implementation of the different migration tasks provided by the
TTC2020 migration case resources \cite{ttc2020resources} load
the different start models from XML files. After the first
migration the task implementation may retrieve the migrated 
model and it may modify the migrated model directly via 
dynamic EMF means. This means, the task implementation does 
not use our editor commands to load or to modify the models. 
Thus, our first task is to \textit{parse} the provided model and 
to derive the editing commands that correspond to it. 

For this purpose, our editors provide a parse method that 
basically uses a visitor to travel through the current model. 
For each model object our visitor then calls special parsePerson
or parseDog methods, respectively. 

Listing~\ref{m2.parse} shows the parsePerson 
method of our M2Editor. Basically, Line~6 to Line~16 of 
Listing~\ref{m2.parse} use dynamic EMF features to retrieve
the parameters needed for the corresponding HavePerson command. 
Line~17 to Line~21 then create the desired HavePerson command and 
provide its parameters. Finally, Line~22 executes the command. 
Thereby, the editor adds the new command to the event store. 

\begin{lstlisting}[language=Java, numbers=left, captionpos=b, 
label={m2.parse}, escapeinside={\%}{\%},
caption={M2Editor::parse()}
]
public class M2Editor 
{
   ...
   private void parsePerson(EObject instance)
   {
      EClass eClass = instance.eClass();
      String name = (String) instance.eGet(eClass.getEStructuralFeature("name"));
      int age = -1;
      EStructuralFeature ageFeature = eClass.getEStructuralFeature("age");
      if (ageFeature != null) {
         age = (Integer) instance.eGet(ageFeature);
      }
      EStructuralFeature ybirthFeature = eClass.getEStructuralFeature("ybirth");
      if (ybirthFeature != null) {
         age = 2020 - (Integer) instance.eGet(ybirthFeature);
      }
      HavePerson havePerson = new HavePerson()
            .setName(name)
            .setAge(age);
      String id = getPersonId(instance);
      havePerson.setId(id);
      execute(havePerson);
   }
   ...
\end{lstlisting}

One crucial step during parsing is the retrieval of ids for the model objects, 
cf. Line~20 of Listing~\ref{m2.parse}. Usually, we require that the model objects 
have an id attribute. In the TTC2020 migration case this is not true. We solve this 
by storing the ids for model objects within the hash tables that are used by our 
getOrCreate methods. To look up the id of a model object that is already stored in 
such a hash table, we search e.g. the hash table for Persons for the 
given instance. If an entry exists, we return the corresponding id. 
If there is no entry yet (i.e. after loading the initial object model) we just
create a new id and add the instance to the hash table under this new id and then return 
the new id. Within the backward migration step the hash table will already contain the 
instance and the old id is retrieved. 

There is one special case when merging the commands created by the parse methods into the 
editor's event store: In migration Task4 of the TTC2020 migration case 
\cite{ttc2020migration}, in model M2 the Dog has no age attribute. Thus, when we parse 
the Dog object during the backward migration step, the parseDog method will not find 
any age information. However, this age information is still contained within the original 
HaveDog command that has been received from M1Editor and that has been executed and added 
to the event store of M2Editor during forward transformation. To keep the age information alive, 
if the parseDog method cannot find the age attribute in the parsed model object, it tries 
to retrieve the old command from the M2Editor event store and copies the age information 
from there into the new command. 

Altogether, our parsing approach utilizes that our command execution is overriding and 
commutative. Due to the commutativity, the parsing visitor may visit the objects of the 
current input model in any order and thus create the editor commands in any order. 
Due to the overriding property, the backward migration may just (re)execute the detected 
commands and this will overwrite the old commands and blend into the event store, 
easily.\footnote{Fortunately, the TTC2020 migration case does not include any delete 
operations during model modification. To handle deletion of model objects one has 
to remove the corresponding old commands from the event store and one has 
to propagate this command removal to the other editor and the other editor needs 
to remove the command, too, and it needs to undo the command in order to roll back 
the corresponding model changes. This rollback needs some careful dealing with our 
getOrCreate operations.}

\section{Results}\label{Results}

Overall, our editor and command approach was very well suited for the TTC2020 
migration case. We were able to address all migration task with the 
same implementation. While our design might appear a little bit over engineered 
we took benefit from the FulibServiceGenerator, a tool we are just building and 
that generates a lot of boiler plate code for editors and commands. After all 
we just had to implement the run methods of the 2 commands for the two editors 
and the parsing methods for the two editors as described above. 

We believe that the use of overwriting and commutative commands provides a great  
leverage for the parsing and merging of command sets. Overall, our design is 
able to handle much more complicated migration cases which we address in our current 
work. 

Unfortunately, our performance is very poor, on a test run the Fulib solution took 
9.6 seconds for 10000 iterations while the original solution used only 0.3 seconds. 
We had no time to go into the details of this. 

You find our solution on: 


Github: https://github.com/fujaba/ttc2020MigrationCaseByFulib

Docker: zuendorf/fulib-solution-ttc2019

\bibliographystyle{alpha}
\bibliography{sekassel.bib}

\end{document}